ISO/IEC JTC 1/SC 29/WG 1
(ITU-T SG16)

# Coding of Still Pictures

**JBIG**
Joint Bi-level Image
Experts Group

**JPEG**
Joint Photographic
Experts Group

| | |
|---|---|
| **TITLE:** | **Technical description of the EPFL submission to the JPEG DNA CfP** |
| **AUTHORS:** | Davi Lazzarotto (davi.nachtigalllazzarotto@epfl.ch)<br>Jorge Encinas Ramos<br>Michela Testolina (michela.testolina@epfl.ch)<br>Touradj Ebrahimi (touradj.ebrahimi@epfl.ch) |
| **PROJECT:** | JPEG DNA |
| **STATUS:** | Final |
| **REQUESTED ACTION:** | For information and feedback |
| **DISTRIBUTION:** | WG1 |

**Contact:**
ISO/IEC JTC 1/SC 29/WG 1 Convener – Prof. Touradj Ebrahimi
EPFL/STI/IEL/GR-EB, Station 11, CH-1015 Lausanne, Switzerland
Tel: +41 21 693 2606, Fax: +41 21 693 7600, E-mail: Touradj.Ebrahimi@epfl.ch

# Table of Contents

## 1. SCOPE

This document provides a technical description of the codec proposed by EPFL to the JPEG DNA Call for Proposals [1]. The codec we refer to as V-DNA for its versatility, enables the encoding of raw images and already compressed JPEG 1 bitstreams, but the underlying algorithm could be used to encode and transcode any kind of data. The codec is composed of two main modules: the image compression module, handled by the state-of-the-art JPEG XL codec, and the DNA encoding module, implemented using a modified Raptor Code implementation following the RU10 (Raptor Unsystematic) [2] description implemented by [3]. The code for encoding and decoding, as well as the objective metrics results, plots and biochemical constraints analysis are available on ISO Documents system with document number WG1M101013-ICQ-EPFL submission to the JPEG DNA CfP.

## 2. WORKFLOW DESCRIPTION

Figure 1 shows the operational flow of the codec. The two displayed inputs, i.e. an uncompressed image or a JPEG 1 image correspond to the *encoding* and to the *transcoding* workflows, respectively. In the first case, JPEG XL is used to compress the input uncompressed image (`.png`) at a given quality level controlled by a quality parameter $q$. In the *transcoding* workflow, the input JPEG 1 bitstream (`.jpg`) is losslessly transcoded to JPEG XL, allowing the reduction of the bitstream in size while maintaining bit-exact reconstruction at the decoder side. In either cases, the JPEG XL bitstream is then served as input to the Raptor encoder, producing a FASTA file (`.fasta`) containing a pool of oligos that verify the biochemical constraints defined in [4].

At the decoder side, the FASTA file is input to the Raptor decoder, which combines the oligos and decodes them to obtain the same JPEG XL bitstream as that generated at the encoder side, without losses. Finally, the bitstream is decoded with the JPEG XL engine to obtain the reconstructed image. Note that in the *transcoding* workflow, the JPEG XL bitstream is decoded into an uncompressed image (`.png`) even if the input to the encoder is a JPEG 1 bitstream.

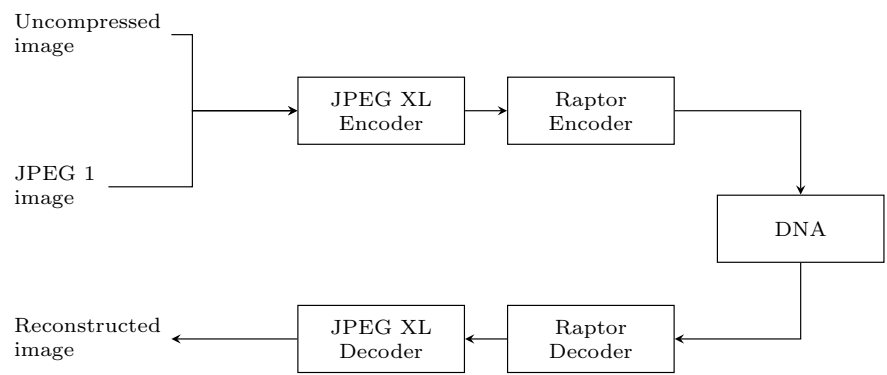

*Figure 1: Codec workflow*

## 3. ENCODER: TECHNICAL DESCRIPTION

Both in the *encoding* and *transcoding* workflows, the compression of the input image is handled by the JPEG XL engine. This algorithm is extensively described in other publications [5] and JPEG documents. In the *encoding* workflow, the JPEG XL compression of the input image is conducted with default parameters, except for the quality parameter $q$ that is adapted to control the nucleotide rate of the encoded image. In the *transcoding* workflow, since the transcoding from the input JPEG 1 bitstream is numerically lossless, there is no quality parameter to be modified, and therefore all default parameters were kept.

The implementation of this proposal uses the JPEG XL engine version 0.8.1, which was obtained from the *libjxl* GitHub repository available at https://github.com/libjxl/libjxl. The *cjxl* and *djxl* binaries were used for compression/transcoding and decompression, respectively.

In both *encoding* and *transcoding* workflows, a JPEG XL bitstream is served as input to the Raptor encoder. Figure 2 illustrates the compression stack, which evidences that the image codec and the Raptor codec operate independently from each other. The remainder of this section describes in details the operation of the Raptor encoding, assuming that a JPEG XL bitstream is served to it as input.

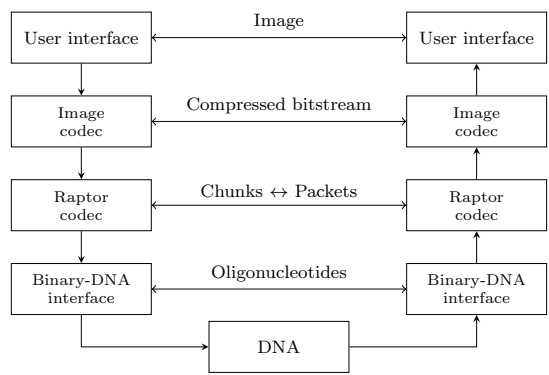

*Figure 2: DNA compression stack*

a. **Payload encoding**

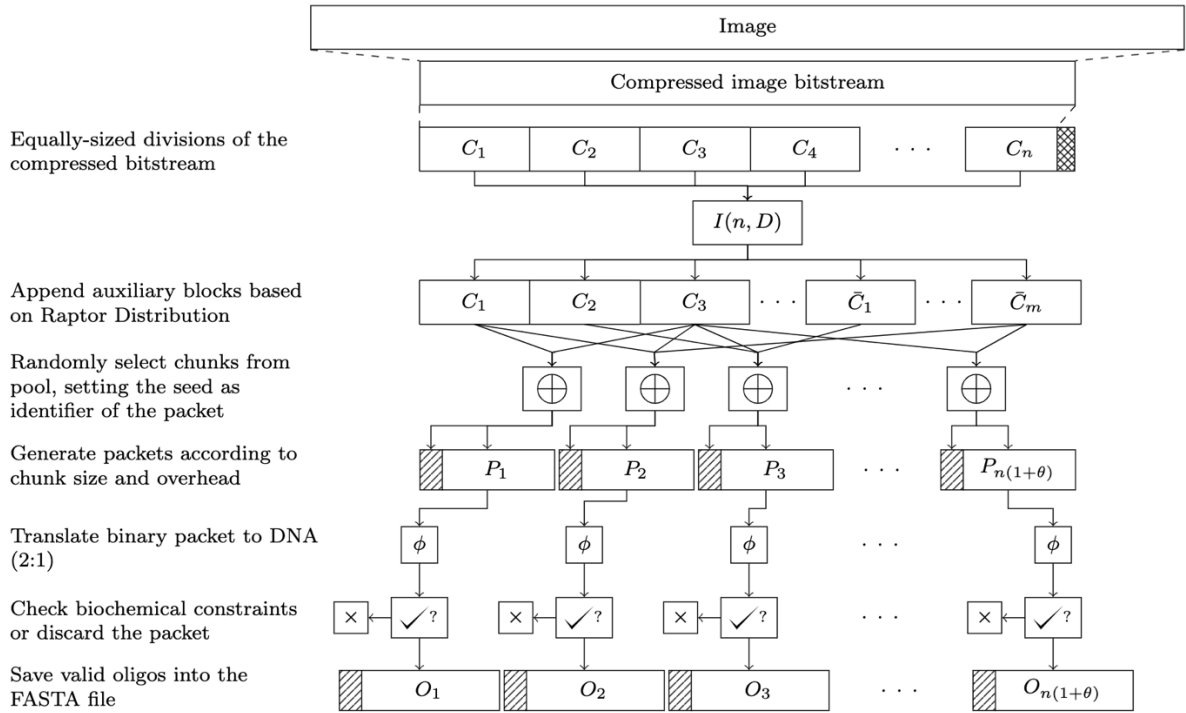

*Figure 3: Oligo generation*

In Figure 3, the algorithm that produces a set of oligos from the input JPEG XL bitstream is illustrated, which is identical for both the *encoding* and *transcoding* workflows. The process begins by dividing the binary input stream in equally sized blocks of information. The size of these blocks relates to the size of the resulting oligos at the output and is here configured to be of 46 bytes, leading to oligos of 200 nucleotides (nts) to comply with the length limitation constraints imposed in the Call for Proposals. If desired, the block size can be easily adjusted to accommodate different oligo sizes. The relationship between block size and oligo size is explained in more details next sections. The last block is padded with 0x00 bytes to keep the same block size as the others. Given a file size **F** in bytes and a chosen block size **c**, the pool of source blocks will contain $n = \lceil F/c \rceil$ blocks.

These **n** blocks are preprocessed in the intermediate block generation function, **I(n, D)**, which corresponds to the first stage of Raptor encoding. Its input parameter **n** is the number of source blocks produced previously, while **D** is the Raptor Distribution that controls the randomness introduced by the algorithm. This distribution is pre-defined and is known by both the encoder and the decoder. **I** produces an additional **m auxiliary blocks** of the same size as the **n** source blocks, generated by Gray encoding and LDPC (Low Density Parity Check) coding. More details on how these intermediary codes are implemented can be found in [3]. The choice for these two codes can be explained due to the combined simplicity and strong correction capabilities, enabling recovery of completely missing oligos.

The second stage of the Raptor encoding is the Luby Transform (LT). This coding phase begins from the previously defined pool of intermediate blocks of size $n + m$, $\{C_0 \ldots, C_n, \bar{C}_1, \ldots, \bar{C}_m\}$. In every round, the LT code selects a given number of blocks (known as degree) according to distribution **D**. The data contained in these blocks is combined by means of exclusive-or (XOR) operations to produce a packet. In other words, each bit of the produced packet is the result of an XOR operation applied over the bits at the same position in all the intermediate blocks used to generate that packet. The identifier of the packet is the seed used to feed the random number generator that is selected the intermediate blocks according to **D** during the packet generation process. This seed needs to be included in the packet as an Id, enabling the decoder to identify it as well as reconstruct the source blocks. The space for Ids (or seeds) is 32 bits, and it is located at the beginning of each binary packet that is LT encoded. It is worth noting that the Ids are not stored in the clear, and they are XORed with a fixed-length predefined mask, before storage, to prevent the Ids from reducing the oligo quality. A structure of these packets can be observed in Figure 4.

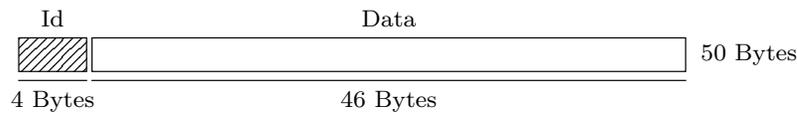

*Figure 4: Payload packet structure*

The 46 bytes of payload coincide with the previously selected block size and consists on the XORed data of the participating blocks. These packets are then translated into DNA using a 2 bits/nucleotide mapping represented by the function:

$$\phi : \{00, 01, 10, 11\} \rightarrow \{A, C, G, T\}$$

Since this translation is not constraint compliant, this process may generate oligos that are not valid. As a consequence, a filtering mechanism is implemented in order to ensure the quality of the oligos included in the final FASTA file:

1. Given a packet **P**$_i$, obtain its DNA structure $O_i = \phi(P_i)$.
2. Verify the oligo's quality verifying the biochemical constraints defined in [4].
   - Homopolymer runs: oligos with homopolymers with length higher than 4 are excluded.
   - Pattern repetitions: any pattern of 3, 4 or 5 nucleotides should never be repeated more than 2 consecutive times. Moreover, any pattern of 6 or 7 nucleotides should never be repeated more than 3 consecutive times.
   - GC content: only oligos with CG content constituting between 40% and 50% of their total nucleotides were accepted.
3. If **O**$_i$ is valid, it is added to the pool of valid oligos. If not, **P**$_i$ and **O**$_i$ are discarded and a new packet is generated via LT coding.

The structure of the generated oligos can be observed in Figure 5.

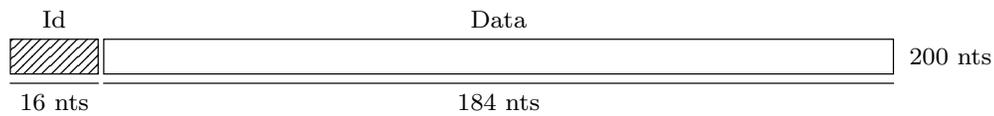

*Figure 5: Payload oligo structure*

This process iterates until $n \cdot (1 + \theta)$ valid oligos are generated, where $\theta$ is the overhead that corresponds to the amount of additional packets that are needed to ensure that the input data can be decoded. While ideally a minimal overhead of 0 would be desired, it is possible due to the randomness of the process that some source blocks are not represented in the generated packets, in which case the decoder would not be able to retrieve their information. For this reason, an overhead greater than 0 is usually required to ensure that all the source blocks are represented in the final oligo pool, allowing for correct decoding. Empirical tests revealed that an overhead of 1.5% is enough to ensure that all the images from the dataset can be decoded. However, the proposed codec provides an alternative to achieve the minimal overhead value for correct decoding by means of a pseudo-decoder that is executed at encoding time and tests if the FASTA can be correctly decoded at each added packet. The results generated for the Call for Proposals were produced using this pseudo-decoder, which increases encoder complexity but attains optimal nucleotide rates for a given quality.

The Raptor codec was implemented using the NOREC4DNA [3] Python library as its baseline, although modifications were introduced to the library in order to enforce it to follow the oligo structure described above.

## b. Header encoding

The only information aside from the oligos themselves that is required to decode is the number of source blocks **n** generated during the encoding. This is required to recreate the auxiliary blocks at the decoder side in order to remove them from the XORed payload of received packets upon reconstruction. Therefore, the value of **n** needs to be transmitted. The solution adopted in this proposal is to include it in a *header segment*, which is appended to the first oligo of the FASTA file.

The header is first formed in the binary domain and then converted to DNA with a coding scheme based on the one defined by *Blawat et al.* in [6]. The binary header contains the following information:

- The total number of source blocks **n**, which is defined as a 32-bit integer.
- The total number of bytes in the last source block, which is used in the decoder side to remove the padding added during encoding. This value is encoded as an 8-bit integer.
- One byte defining the mode and extension (0x00 for *encoding*, 0x11 for *transcoding* from JPEG 1 bitstreams).
- One byte to indicate the *S* parameter, which is used for decoding the header as described in the following paragraphs.

The code used to encode the header [6] avoids homopolymer runs above 3 nucleotides and maintains balanced GC content by using a dictionary to encode each byte into 5 nucleotides, achieving a conversion rate of 1.6 bits/nucleotide. However, even if the occurrence of patterns is uncommon, their absence cannot be guaranteed by this coding scheme. For this reason, a header transformation mechanism has been implemented, where all the bytes of the binary header, except for the last one, are XORed with a random sequence of 6 bytes. This sequence is generated after feeding the $S$ parameter as a seed to a random number generator. The last byte of the transformed binary header is left untouched to extract the $S$ value in decoding.

The transformed binary header is then converted to DNA using the mechanism defined in [6], and is then tested for the biochemical constraints. If the nucleotide sequence fails the test, a new $S$ value is produced to feed the random number generator, which is then used to produce a new transformed binary header which in its turn is used to produce a new nucleotide sequence. This process is repeated iteratively until a valid DNA sequence is found, with the initial value of $S$ starting at 0 and being able to go up to 255.

When a valid *header segment* is found, it is appended to the first oligo of the FASTA file. In order to avoid the generation of a homopolymer between the end of the first oligo and the beginning of the *header segment*, another nucleotide is included between them, which is forced to be different to the last nucleotide of the first oligo and the first nucleotide of the *header segment*.

### c. Error-correcting capabilities

The encoder implementation described above allows natively for error-correction coming from its Raptor codec. Although a minimal oligo overhead $\theta$ was adopted to produce the results for the Call for Proposals to reduce the nucleotide rate, purposefully adopting larger overheads would allow to recover the original information even if oligos were completely lost at the DNA channel, as long as a minimal amount of oligos were correctly retrieved from the pool.

This feature is however not enough to correct errors generated inside an oligo. Fortunately, the NOREC4DNA library includes an implementation of Reed-Solomon codes for this purpose, which would allow the codec to correct a certain amount of substitution errors inside the oligos. Although this feature was not tested in the proposal, its inclusion can be easily evaluated in future core experiments.

## 4. DECODER: TECHNICAL DESCRIPTION

The decoding flow begins from a recovered FASTA, from which the header segment is first extracted. For that, the oligo with a larger size than the remaining is identified, and the appended portion is obtained. The header is parsed, firstly extracting the $S$ parameter used to decode the rest of the header after reproducing the transformation done in the encoder. The remaining information from the header is obtained afterward, including the number of source blocks **n**.

Once **n** is obtained, it is passed, along with the rest of the FASTA file to the Raptor decoder. The first step is to translate the nucleotides of each oligo back into their binary counterparts using $P_i = \phi^{-1}(O_i)$. Following that, recreation of the **m** auxiliary blocks using **n** takes place. These are needed to XOR them with other packets as well as later removing them to return the original pool of **n** source blocks (without any auxiliary information). After obtaining all blocks and removing the auxiliary information, the decoding can be reduced and solved by means of Gaussian elimination with partial pivoting (GEPP), which enables uniform and fast decoding. Although the GEPP method may be less advantageous than other approaches such as those based on Belief Propagation Algorithms for large file sizes, it was here retained because it was completely implemented in the source library. However, more sophisticated decoding may be explored in future to aim for complexity reduction.

The resulting bitstream is later processed according to the *mode* and *extension* defined in the header, which sets the codec to either *transcoding* or *encoding* mode. The JPEG XL engine is finally used to reconstruct the image.

## 5. EXPERIMENTAL CODEC

The proposed codec separates the image compression task from the translation to nucleotides. For this reason, any compression engine could be used to generate a bitstream from the input uncompressed image. While the JPEG XL standard allows to achieve high rate-distortion performance, JPEG AI has recently demonstrated excellent performance. Even if JPEG AI was not formally included in the proposal due to its early development stage in the standardization process, its use is explored as an experimental codec. In particular, the JPEG AI VM 4.1 was experimentally used instead of JPEG XL in the V-DNA framework and was employed to compress the images of the dataset for the *encoding* workflow at five different bitrates following the configuration files contained in the reference software (tools off). These results are also reported in this document as an exploration study to provide insights for the potential improvements that could be brought to the proposed codec.

## 6. RESULTS

### a. Encoding workflow

The *encoding* workflow was used to compress the uncompressed images of the JPEG DNA dataset at nucleotide rates as close as possible to the target rates defined in the Call for Proposals. For that purpose, the $q$ parameter used for the JPEG XL compression was the only parameter to be tuned, since the remaining parts of the encoder translate binary data to DNA at a fixed rate. The corresponding $q$ values employed at each target rate and each image are provided in the Table 1.

| Image | r1 | r2 | r3 | r4 | r5 | r6 | r7 |
|---|---|---|---|---|---|---|---|
| **00001_1192x832** | 96 | 94 | 92 | 90 | 86 | 81 | 76 |
| **00002_853x945** | 95 | 94 | 92 | 90 | 88 | 84 | 81 |
| **00003_945x840** | 96 | 95 | 92 | 89 | 85 | 78 | 70 |
| **00004_2000x2496** | 91 | 88 | 84 | 79 | 74 | 65 | 62 |
| **00005_560x888** | 94 | 92 | 89 | 86 | 82 | 76 | 69 |
| **00006_2048x1536** | 97 | 96 | 94 | 92 | 89 | 85 | 80 |
| **00007_1600x1200** | 96 | 94 | 92 | 88 | 84 | 78 | 70 |
| **00008_1430x1834** | 98 | 96 | 94 | 92 | 88 | 84 | 78 |
| **00009_2048x1536** | 99 | 98 | 96 | 94 | 91 | 87 | 80 |
| **00010_2592x1946** | 95 | 94 | 92 | 90 | 87 | 83 | 78 |

*Table 1: quality parameter used for each image at each target rate*

The objective metrics defined in the JPEG DNA Common Test Conditions were computed on the decoded images. The rate-distortion plots from Figure 6 contain the average metric values across all images of the dataset for each rate, together with 95% confidence intervals. These values are plotted against those provided with the Call for Proposals for Anchor 2, which corresponds to the sole anchor in the *encoder* category. Moreover, the other two *encoder* proposals submitted by other proponents to the JPEG DNA Call for Proposals, *HiDNA* and *BioCoder*, were included in the plots, using the metric values extracted from the proposal package in their submission. The rate-distortion plot for each image in the JPEG DNA dataset is available in the submission package (WG1M101013-ICQ-EPFL submission to the JPEG DNA CfP).

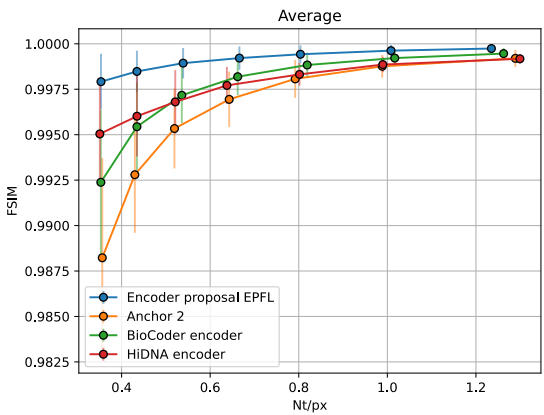
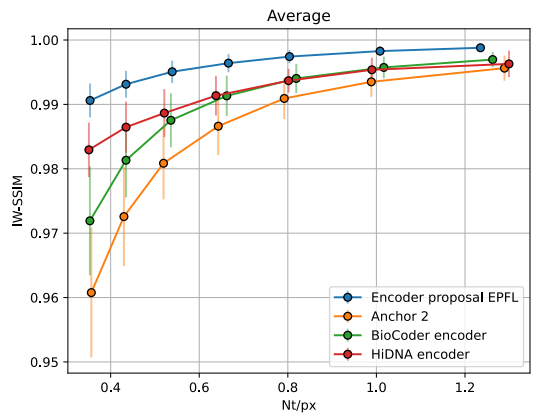

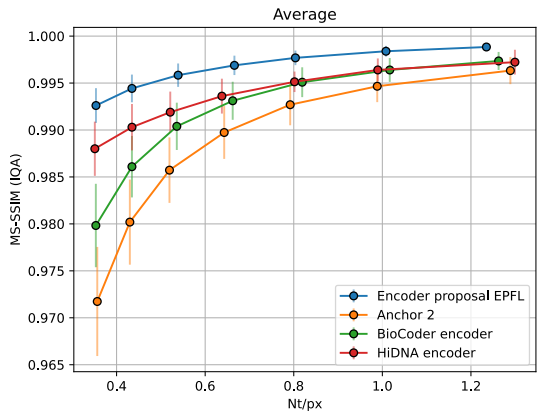

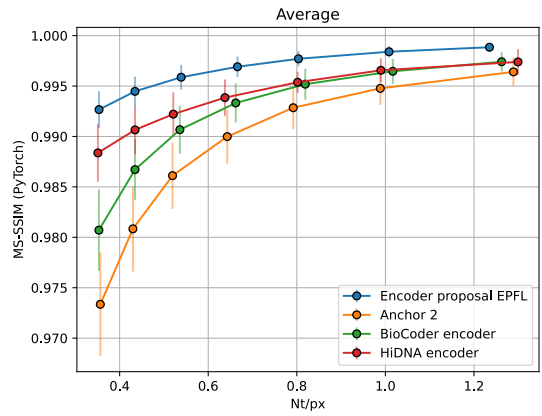

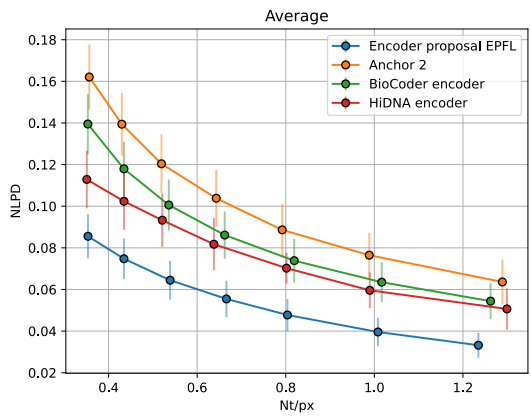

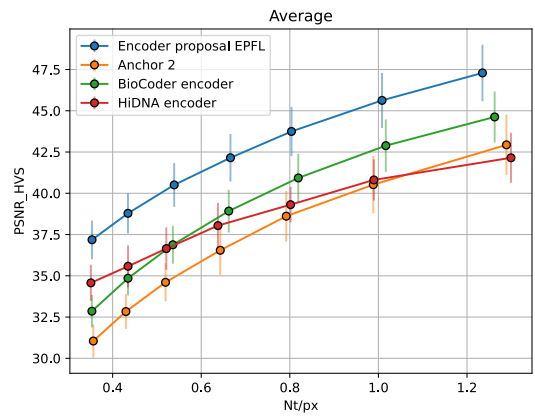

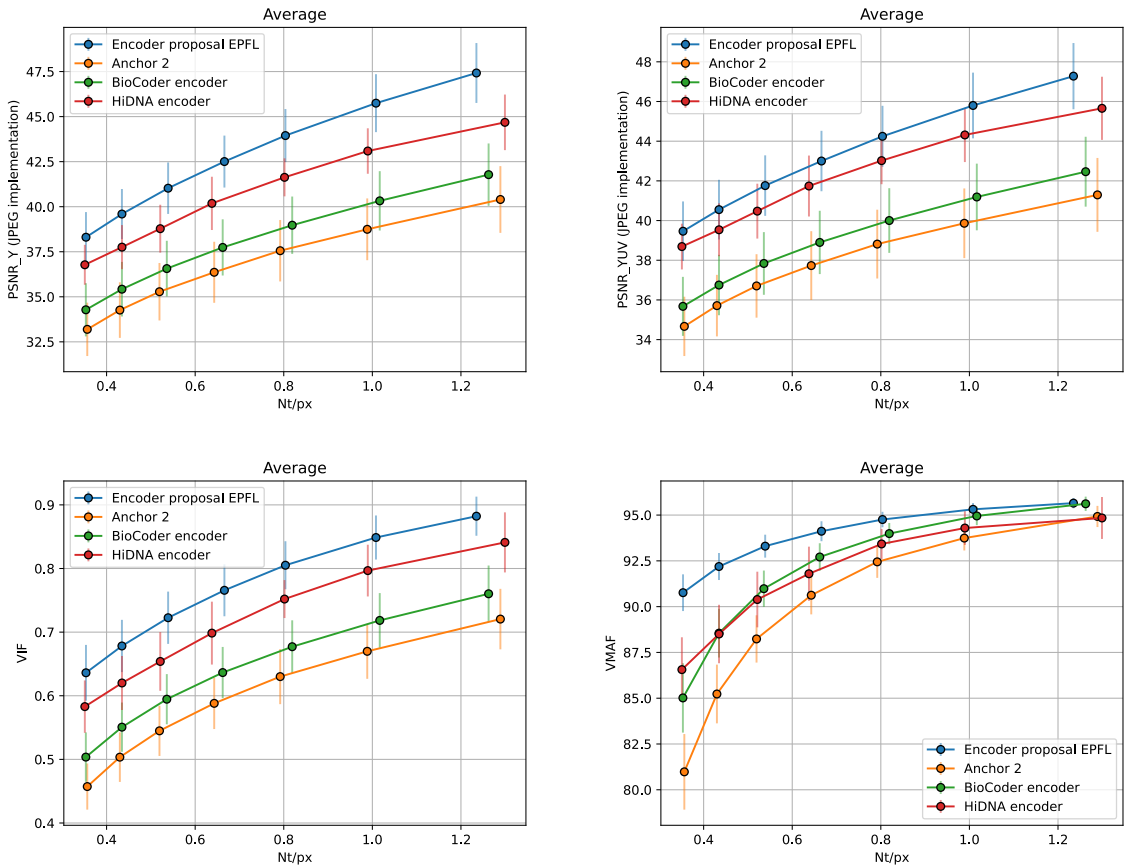

*Figure 6: Rate-distortion plots for encoding workflow of the proposal compared anchor and other proposals*

The presented plots demonstrate that the EPFL proposal outperforms both the anchor and the other proposals according to all the metrics, at all target nucleotide rates. This effect is observed with higher intensity for some metrics, such as PSNR-HVS-M, and to a lesser extent for others, such as PSNR YUV. However, in all cases, the average metric values are higher than the second best performing curve, i.e. the *HiDNA* encoder.

### b. Transcoding workflow

The *transcoding* workflow was used to transcode all the JPEG 1 bitstreams from the JPEG DNA dataset, i.e. 10 files for each source image, each corresponding to a different bitrate. Since JPEG XL transcoding from JPEG 1 bitstreams is lossless, it is not possible to control the nucleotide rate at the transcoding step. For this reason, the rates of the transcoded files do not correspond to the target nucleotide rates. If rate control of transcoded files is desired, then the *q* parameter used during JPEG 1 compression should be adapted, which was not possible in this case since the JPEG 1 bitstreams are directly provided in the JPEG DNA dataset for fairness and to avoid interference from the performance of the exact JPEG 1 encoder implementation.

The rate-distortion plots of the *transcoding* workflow averaged over all the images in the JPEG DNA dataset are compared with Anchor 1 and Anchor 3, which also belong in the *transcoder* category, as well as the *BioCoder* proposal, and can be observed in Figure 7. The rate-distortion plot for each image in the JPEG DNA dataset is available in the submission package (WG1M101013-ICQ-EPFL submission to the JPEG DNA CfP).

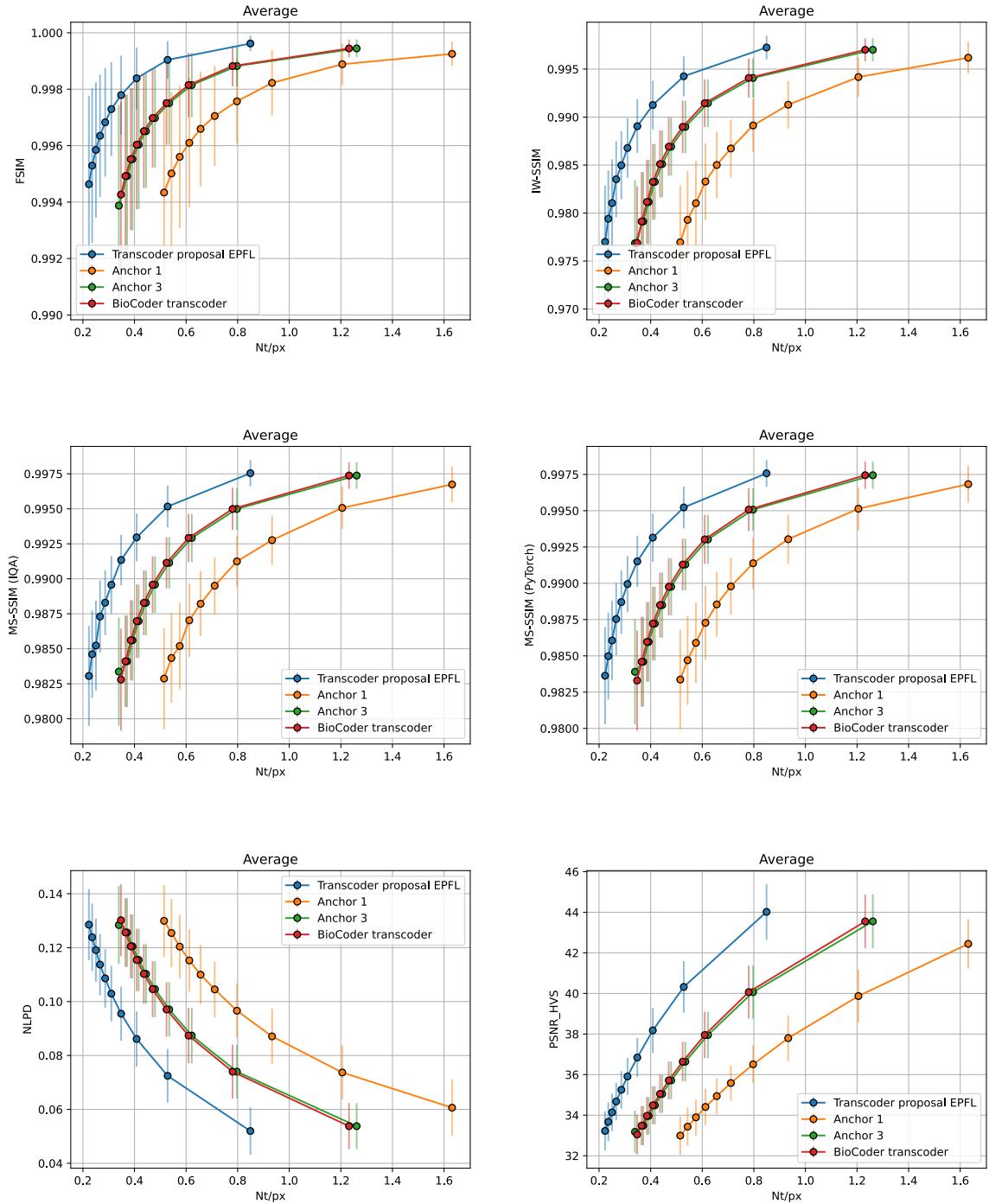

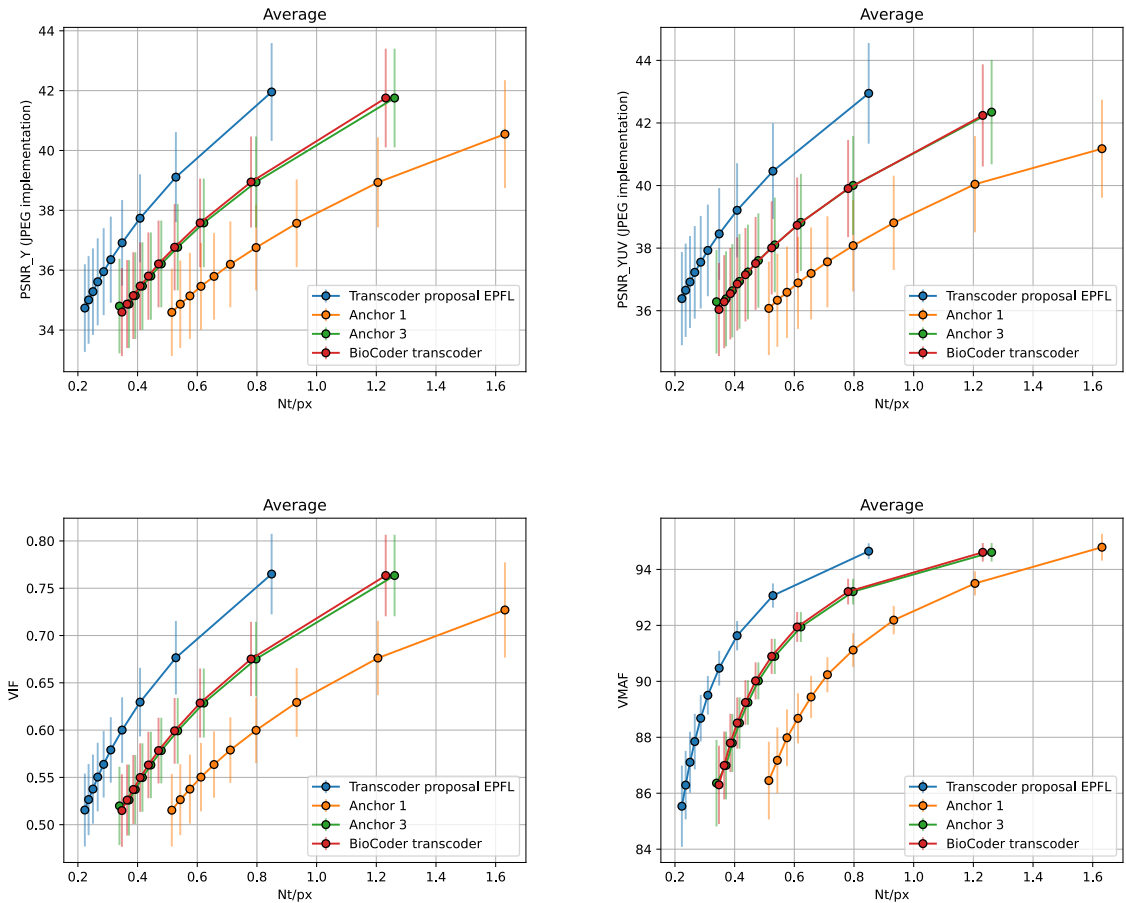

*Figure 7: Rate-distortion plots for transcoding workflow of the proposal compared anchors and other proposals*

The presented plots show that the average metric values are nearly the same for the EPFL proposal, the anchors, as well as for the *BioCoder* transcoder, which is expected since they are all lossless transcoders. The small observed differences are likely due to small differences in the decoder implementation and metrics computation. However, the nucleotide rates of all the transcoded FASTA's are lower for this proposal, demonstrating its effectiveness.

## c. Experimental codec

As detailed in Section 5, an experimental codec was also tested where JPEG XL was replaced by JPEG AI in the *encoding* workflow. Since the target bitrates defined in the configuration files of the JPEG AI VM 4.1 do not allow for a broad selection of rates, it was not possible to achieve results following the target rates imposed by the Call for Proposals. However, the obtained results are reported here as a demonstration of the versatility of this proposal, as well as an indicator of a potential performance increase that could be brought if JPEG XL is replaced by JPEG AI. The rate-distortion plots for the metrics PSNR Y and MS-SSIM are presented in Figure 8.

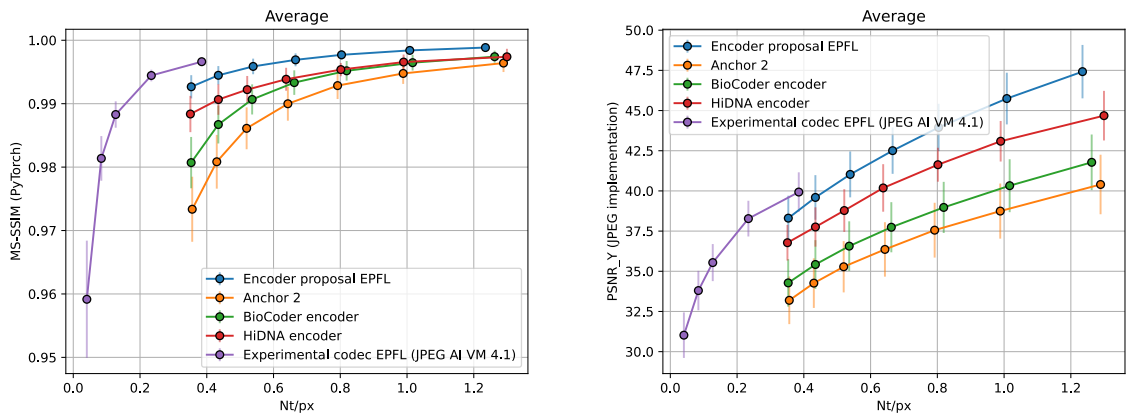

*Figure 8: Rate-distortion plots for the experimental codec compared anchors and other proposals*

The plots demonstrate that the experimental codec allows to achieve superior metric values when comparing its highest rate point to the lowest rate point of the Call for Proposals. However, this codec is not able to achieve the highest qualities with its default configurations. This analysis suggests that tuning the compression parameters of JPEG AI could allow for better performance than the current proposal if the observed behavior is maintained for higher bitrates. Nevertheless, in its current configuration, this experimental codec is not able to achieve the quality levels required by the use cases requiring DNA-based storage.

## 7. BIOCHEMICAL CONSTRAINTS COMPLIANCE

As described in Section 3.a, the proposed encoder guarantees compliance with all the biochemical constraints in [4]. Such assumption is verified through the software provided as part of the JPEG DNA Metrics package.

The FASTA files obtained with both the *encoding* and *transcoding* workflow have been inspected using the provided software. The analysis revealed the following:

- <u>Strand length limitations</u>: all oligos in the analyzed FASTA files have length of 200 nucleotides, with exception of the first oligo which has a size of 236 nucleotides due to the added *header segment*.
- <u>Homopolymer runs</u>: no homopolymer of size 4 or larger have been detected in the analyzed FASTA files.
- <u>GC content balance</u>: all the oligos in the analyzed FASTA files shows a compliant percentage of GC content, namely all oligos have GC content between 40% and 60%.
- <u>Repetition of patterns</u>: no patterns have been detected in the analyzed FASTA files.

## 8. COMPUTATIONAL COMPLEXITY

As an indication of computational complexity, the times to encode and decode each image and bitstream were measured both in the *encoding* and *transcoding* workflows. The computation was conducted on a platform with a Intel(R) Core(TM) i9-11900K @ 3.50GHz CPU running Ubuntu 20.04.4.

### a. Encoding workflow

The minimum, maximum and average encoding and decoding times over all the images and rates for the *encoding* workflow across the entire dataset are reported in Table 2.

|  | Min | Max | Average |
|---|---|---|---|
| Encoding time (s) | 2.32 | 1128.62 | 132.67 |
| Decoding time (s) | 0.75 | 956.26 | 104.29 |

*Table 2: encoding and decoding times of the encoding workflow for the entire dataset*

Encoding time is consistently higher than decoding, likely due to the fact that a pseudo-decoder is already implemented at the encoder side and is executed for each added packet. The average encoding and decoding times lie approximately around two minutes. However, the range between minimum and maximum times is large. For that reason, an analysis separating the time values by source image and rate is also conducted and reported in Tables 3 and 4.

| Rate | Encoding time (s) | | | Decoding time (s) | | |
|---|---|---|---|---|---|---|
|  | Min | Max | Average | Min | Max | Average |
| r1 | 12.04 | 1128.62 | 380.41 | 4.78 | 956.26 | 315.11 |
| r2 | 9.18 | 612.55 | 226.67 | 3.40 | 496.32 | 179.99 |
| r3 | 6.48 | 272.16 | 127.12 | 2.01 | 224.73 | 98.56 |
| r4 | 4.90 | 200.05 | 85.36 | 1.54 | 158.87 | 64.32 |
| r5 | 3.81 | 130.04 | 52.35 | 1.19 | 98.97 | 36.84 |
| r6 | 3.11 | 88.83 | 33.41 | 0.91 | 64.90 | 21.54 |
| r7 | 2.32 | 61.07 | 23.37 | 0.75 | 40.66 | 13.68 |

*Table 3: encoding and decoding times of the encoding workflow separated by rate*

| Image | Encoding time (s) | | | Decoding time (s) | | |
|---|---|---|---|---|---|---|
| | Min | Max | Average | Min | Max | Average |
| 00001_1192x832 | 11.49 | 63.89 | 30.36 | 4.89 | 43.19 | 17.92 |
| 00002_853x945 | 3.92 | 12.04 | 7.20 | 1.17 | 4.78 | 2.62 |
| 00003_945x840 | 2.32 | 12.49 | 6.15 | 0.75 | 5.05 | 2.17 |
| 00004_2000x2496 | 40.21 | 435.49 | 165.48 | 23.94 | 381.00 | 131.76 |
| 00005_560x888 | 2.69 | 12.25 | 6.22 | 0.78 | 5.15 | 2.22 |
| 00006_2048x1536 | 38.43 | 504.07 | 199.04 | 23.89 | 404.51 | 152.90 |
| 00007_1600x1200 | 11.14 | 90.50 | 38.33 | 4.47 | 64.20 | 24.02 |
| 00008_1430x1834 | 36.77 | 960.09 | 289.33 | 22.24 | 823.16 | 239.12 |
| 00009_2048x1536 | 25.64 | 1128.62 | 329.95 | 14.06 | 956.26 | 268.99 |
| 00010_2592x1946 | 61.07 | 584.64 | 254.63 | 40.66 | 463.75 | 201.17 |

*Table 4: encoding and decoding times of the encoding workflow separated by source image*

The compression rate is found to play a large role in the execution time, with lower rates naturally being faster to encode and decode. Likewise, input images with lower resolution usually take shorter time to encode and decode. These values are directly related to the number of source blocks that are needed to represent the compressed JPEG XL bitstream, which increase the time taken by the Raptor Coder.

### b. Transcoding workflow

Similarly to the *encoding* workflow, the compression and decompression times were measured for the *transcoding* workflow, and is reported for the whole dataset as well as separately for each rate and each image in Tables 5, 6 and 7.

| | Min | Max | Average |
|---|---|---|---|
| Encoding time (s) | 1.53 | 476.89 | 34.06 |
| Decoding time (s) | 0.55 | 391.27 | 23.24 |

*Table 5: encoding and decoding times of the transcoding workflow for the entire dataset*

| Rate | Encoding time (s) | | | Decoding time (s) | | |
|------|------|------|---------|------|------|---------|
|      | Min  | Max  | Average | Min  | Max  | Average |
| r1   | 7.17 | 476.89 | 154.19 | 2.55 | 391.27 | 122.07 |
| r2   | 4.15 | 111.54 | 51.08  | 1.25 | 84.18  | 35.49  |
| r3   | 3.13 | 60.83  | 29.77  | 0.92 | 43.32  | 18.33  |
| r4   | 2.43 | 46.71  | 22.56  | 0.76 | 31.03  | 13.14  |
| r5   | 2.39 | 42.58  | 18.00  | 0.72 | 26.85  | 10.16  |
| r6   | 2.13 | 38.01  | 15.95  | 0.68 | 23.26  | 8.63   |
| r7   | 1.92 | 34.00  | 14.00  | 0.66 | 20.60  | 7.18   |
| r8   | 1.86 | 32.77  | 12.80  | 0.62 | 19.68  | 6.54   |
| r9   | 1.68 | 30.41  | 11.62  | 0.58 | 17.71  | 5.71   |
| r10  | 1.53 | 27.98  | 10.68  | 0.55 | 16.18  | 5.17   |

*Table 6: encoding and decoding times of the transcoding workflow separated by rate*

| Image | Encoding time (s) | | | Decoding time (s) | | |
|-------|------|------|---------|------|------|---------|
|       | Min  | Max  | Average | Min  | Max  | Average |
| 00001_1192x832   | 6.60  | 31.02  | 11.57 | 2.21  | 18.03  | 5.14  |
| 00002_853x945    | 2.27  | 7.17   | 3.35  | 0.77  | 2.55   | 1.10  |
| 00003_945x840    | 1.53  | 7.25   | 2.85  | 0.55  | 2.59   | 0.93  |
| 00004_2000x2496  | 15.73 | 146.77 | 43.37 | 8.06  | 109.48 | 27.66 |
| 00005_560x888    | 1.74  | 7.32   | 2.97  | 0.59  | 2.60   | 0.95  |
| 00006_2048x1536  | 17.31 | 217.48 | 52.08 | 8.31  | 177.61 | 36.32 |
| 00007_1600x1200  | 6.09  | 45.44  | 14.06 | 2.07  | 27.01  | 6.83  |
| 00008_1430x1834  | 16.16 | 314.31 | 66.37 | 7.96  | 268.29 | 49.23 |
| 00009_2048x1536  | 11.36 | 476.89 | 74.24 | 5.03  | 391.27 | 55.22 |
| 00010_2592x1946  | 27.98 | 288.21 | 69.80 | 16.18 | 221.31 | 49.04 |

*Table 7: encoding and decoding times of the transcoding workflow separated by source image*

Again, the encoding times are found to be higher than the decoding times. Moreover, the times are considerably shorter than those for the *encoding* mode, due to the fact that the observed nucleotide rates are lower.

## 9. CONCLUSIONS

The proposed V-DNA codec proposal obtains superior performance compared to anchors both in terms of rate and objective metrics, for encoding raw images as well as for transcoding JPEG 1 bitstreams. The computational complexity cost is not negligible but could be reduced at the encoder side by avoiding the use of the pseudo-decoder and applying a fixed pre-defined overhead. When decoding, other algorithms than the Gaussian elimination with partial pivoting could reduce decoding time for larger files and could be explored in future studies.

V-DNA includes the possibility of error correction, mainly by the Raptor codec which is capable of producing overhead oligos that allow to properly decode the input image even if entire oligos are lost in the DNA channel. Moreover, the library employed to run these codes includes the possibility of adding FEC (Forward Error Correction) codes at the packet level which would further increase the codec's resilience to the high channel distortions produced by operations in the DNA space.

The codec also enables customization of parameters such as the block size or the overhead and includes a prototypical header that can be given a wide range of uses apart from the required signaling explained in this document. Moreover, even if this Call for Proposals is focused on images, the algorithm leveraged in this codec could be applied to other forms of media such as video, plenoptic image modalities, or more generally to any kind of information represented in binary form. This versatility also extends to the biochemical constraints, which could be easily updated in the codec if more accurate models for the DNA error channel are proposed.